\documentclass{article}
\usepackage[utf8]{inputenc}
\usepackage{authblk}
\usepackage[dvips]{graphicx}
\graphicspath{{noiseimages/}}
\usepackage{xcolor}
\usepackage[charter]{mathdesign}
\let\mathcal\undefined
\DeclareMathAlphabet{\mathcal}{OMS}{cmsy}{m}{n}
\usepackage[T2A]{fontenc}
\usepackage{tikz}
\usepackage{empheq}

\usepackage{MnSymbol}
\usepackage{pgfplots}
\pgfplotsset{compat=1.7}
\usetikzlibrary{intersections, pgfplots.fillbetween}
\usetikzlibrary{decorations.pathmorphing}
\usetikzlibrary{arrows.meta}
\usepackage[mode=buildnew]{standalone}
\usepackage[toc,page]{appendix}
\usepackage{amsmath}
\usepackage{eucal}
\usepackage{float}
\usepackage{comment}
\usepackage{bm}
\usepackage{a4wide}
\usepackage{bm}

\usepackage[english]{babel}
\usepackage{hyperref}
\usepackage{caption}
\AtBeginDocument{}
\definecolor{refcolor}{HTML}{CD2600}
\definecolor{tablecolor}{HTML}{373641} 
\definecolor{urlcolor}{HTML}{E12900} 
\hypersetup{pdfstartview=FitH,  linkcolor=refcolor,urlcolor=urlcolor, colorlinks=true,citecolor=refcolor}
\usepackage[page]{appendix}

\newcommand{\w}{\omega}

\author[1,2]{D.V.Diakonov\footnote{\tt dmitrii.dyakonov@phystech.edu}}

\affil[1]{Moscow Institute of Physics and Technology, 141700, Dolgoprudny, Russia}
\affil[2]{National Research Centre “Kurchatov Institute”, 123182, Moscow, Russia
}
\title{\textcolor{black}{Is the Euclidean path integral always equal to the thermal partition function?}}
\begin{document}

\numberwithin{equation}{section}

\maketitle

 The Euclidean path integral is compared to the thermal (canonical) partition function in curved static space-times. It is shown that if spatial sections are non-compact and there is no Killing horizon, the logarithms of these two quantities differ only by a term proportional to the inverse temperature, that arises from the vacuum energy. When spatial sections are bordered by Killing horizons the Euclidean path integral is not equal to the thermal partition function. It is shown that the expression for the Euclidean path integral depends on which integral is taken first: over coordinates or over momenta. In the first case the Euclidean path integral depends on the scattering phase shift of the mode and it is UV diverge. In the second case it is the total derivative and diverge on the horizon. Furthermore we demonstrate that there are three different definitions of the energy, and the derivative with respect to the inverse temperature of the Euclidean path integral does not give the value of any of these three types of energy. We also propose the new method of computation of the Euclidean path integral that gives the correct equality between the Euclidean path integral and thermal partition function for non-compact spaces with and without Killing horizon.

\newpage 
{
  \hypersetup{linkcolor=tablecolor}
  \tableofcontents
}
\newpage

\section{Introduction}

Quantum thermal effects in the gravitational field can be described by two approaches. 

The first approach used e.g. by Gibbons and Hawking \cite{Gibbons:1976ue,Hawking:1976ja} is the Euclidian path integral: 
\begin{align}
    Z^E=\int d[\varphi] e^{-S[\varphi]},
\end{align}
where the integral is expressed in terms of periodic fields in Euclidean time $\tau$ with the period equal to the inverse temperature $\beta$. The advantage of this definition of the partition function is that it is manifestly covariant \cite{Toms:1986sh}.

The second approach is to define the partition function of the canonical ensemble as follows: 
\begin{align}
    Z^C=Tr[e^{-\beta :\hat{H}:}],
\end{align}
where $:\hat{H}:$ is the standard normal ordered Hamiltonian. The advantage of this definition is that it is in accordance with the unitary evolution of the system, but it is not manifestly covariant. 

However, the problem is that the Euclidean path integral $Z^E$ is not always equal to the thermal partition function of the canonical ensemble $Z^C$. In \cite{Allen:1986qi} it is shown that in ulta-static manifold ($g_{00}=1$), Euclidean path integral is equal to the thermal partition function up to the vacuum-energy contribution, which does not depend on temperature. The last term simply shifts the free energy by a constant, therefore it does not affect thermodynamic quantities; for example, it leaves the entropy unchanged. Thus, we will assume that these two approaches are equivalent if they are equal up to a temperature independent shift of the free energy:
\begin{align}
    \log{ Z^E}=\log Z^C-\beta E_0.
\end{align}

But in \cite{Moretti:1997qn}, it is shown that the fundamental statistical-mechanical relation for the Euclidean path integral
\begin{align}
\label{FSR}
    -\partial_\beta \log Z^E = Tr\left[\frac{e^{-\beta :\hat{H}:}}{Z^C} :\hat{H}:\right] ,
\end{align}
does not hold in the Rindler coordinates for non-minimally coupled scalar field. Since the fundamental statistical-mechanical relation obviously holds for the thermal partition function, it means that two approaches are not equivalent in the Rindler coordinates. 

So the aim of this article is to discuss whether the Euclidean path integral $Z^E$ is equal to the thermal partition function of the canonical ensemble $Z^C$ in various cases in curved space-times?

Here we describe what physical situations we have considered in this article. We consider massive non-minimaly coupled scalar field theory defined by the action:
\begin{align}
\label{S}
    S[\varphi]=-\frac{1}{2}\int d^4 x \sqrt{g}\Big[\partial_\mu \varphi(x)\partial^\mu \varphi(x)+m^2 \varphi^2(x)+\xi R \varphi^2(x) \Big],
\end{align}
in a static background metric of the form: 
\begin{align}
    ds^2=-g_{00}dt^2+g_{i j} dx^i dx^j.
\end{align}
The metric depends only on the spatial coordinates, namely $\partial_0 g_{\mu \nu}=0$ and $g_{0 i}=0$. So one can perform the Wick rotation $t\to -i \tau$ to obtain the Euclidean metric: 
\begin{align}
    ds_E^2=g_{00}d\tau^2+g_{i j} dx^i dx^j.
\end{align}
Moreover, if $g_{00} = 1$ the metric is called optical or ultrastatic.

There are three different physical situations, namely when the spatial sections are compact, when spatial sections are non-compact and the manifold is asymptotically flat, and when the spatial sections are non-compact and bordered by Killing horizons.

The paper is organized as follows. In Section 2, we give general overview of methods for calculating the Euclidean path integral. Then we discuss two different representation of the partition function in term of the Euclidean path integral. 
At the end of Section 2, we give a simple proof of the equality between the Euclidean path integral and thermal partition function approaches.  In Section 3, we give another proof of the equality for compact manifold and for non-compact manifold without Killing horizon. At the end of Section 3, we present a new method that gives the correct equality for both compact and non-compact cases with and without Killing horizon. In Section 4, we discuss space-time with Killing horizon. We show that there are three definition of the thermodynamic energy and that they differ by a surface terms. Then we show that the standard method does not give equality between Euclidean path integral and thermal partition function approaches, unlike the new method.

\section{ Euclidean path integral and thermal partition
function\label{sec2}}

In this section, we briefly discuss various methods of calculation of the Euclidean path integral. Then we show that the thermal partition function can be represented in terms of the path integral with the standard action, but with a non-covariant measure, or in terms of Euclidean path integral with the covariant measure, but in terms of an optical metric that is conformaly related to the original one. This allows us to give a simple proof, in the cases of compact and non-compact manifolds without Killing horizon, the equivalence between the Euclidean path integral and thermal partition function approaches holds. But the Euclidean path integral approach does not provide the correct answer for the case of a non-compact manifold with Killing horizon. We will discuss this later cases in the next section.

\subsection{Euclidean path integral }

The Euclidean path integral is defined as follows:
 \begin{align}
    Z^E=\int d[\varphi] e^{-S[\varphi]}.
\end{align}
The fields are assumed to be periodic in Euclidean time with the period $\beta$.  The advantage of this method is that it is manifestly covariant. The functional measure is formally given by \cite{Hawking:1975vcx, Fujikawa:1980vr, Toms:1986sh}:
\begin{align}
\label{measure}
    d[\varphi]=\prod_x \frac{d\varphi(x)}{\sqrt{2\pi}}g^{\frac{1}{4}}=\prod_i \frac{d c_i}{\sqrt{2\pi}}.
\end{align}
Here $c_i$ are the Fourier coefficients of the field in the basis of functions, $\varphi(x)=\sum_i c_i \phi_i(x)$, which are eigenfunction of the Klein-Gordon equation following from \eqref{S}:
\begin{align}
     \left(-\Box_E+m^2+\xi R \right)\phi_i(x)=\lambda_i \phi_i(x),
\end{align}
where the superscript $E$ is standing for the Euclidean signature. They are orthonormal with respect to the following inner product:
\begin{align}
\label{innerq}
    \langle\phi_i| \phi_j\rangle= \int d^4x \sqrt{g} \phi^*_i(x) \phi_j(x)=\delta_{ij}.
\end{align}
The action in terms of the inner product takes the form: 
\begin{align}
    S[\varphi]=\frac{1}{2} \int d^4x \sqrt{g} \varphi(x) \left(-\Box_E+m^2+\xi R \right) \varphi(x)=\frac{1}{2}\langle\varphi| \left(-\Box_E+m^2+\xi R \right) |\varphi\rangle. 
\end{align}
Therefore, one can express the path integral in terms of the functional determinant:
\begin{align}
\label{detpath}
    \int d[\varphi] e^{-S[\varphi]}= det^{- \frac{1}{2}}\left[\frac{-\Box_E+m^2+\xi R }{\mu^2}\right],
\end{align}
where $\mu$ is the normalization scale. 

Note that the functional measure \eqref{measure} is determined by requiring that there is no Einstain anomaly, i.e. it is invariant under general coordinate transformation. 

Let us now consider the Hamiltonian formulation of the Feynman path integral: 
\begin{align}
    Z=\int \prod_x d\varphi(x) d \pi(x) e^{ i \int d^4 x \left[ \pi(x)\partial_0 \varphi(x)-H(\varphi,\pi)\right]  }
    \end{align}
where the canonical momenta defined by: 
\begin{gather}
    \pi(x)=\sqrt{g} g^{00}\partial_0 \varphi(x). 
\end{gather}
We can directly perform the functional integration over canonical momenta: 
\begin{align}
\label{zzz}
    Z=\int \prod_x d\varphi(x) g^{\frac{1}{4}}(g^{00})^{\frac{1}{2}}   e^{ i S[\varphi] }.
    \end{align}
Therefore the functional measure explicitly depend on $g^{00}$. In fact in \cite{Fradkin:1973wke} argued that the noncovariance of the $g^{00}$ factor in the measure is required in order to cancel the leading one-loop divergence, this point was discussed in \cite{Unz:1985wq,Hatsuda:1989qy}. In the next section we will explicitly show that the partition function can be rewritten in terms of a functional integral with the same measure as in \eqref{zzz}.   

The functional measure may be rewritten in the form: 
\begin{align}
    \prod_x d\varphi(x) g^{\frac{1}{4}}(g^{00})^{\frac{1}{2}}=\prod_x d\varphi(x) g^{\frac{1}{4}} e^{\frac{1}{2} Tr \left[ \log \left( g^{00}\right) \right]},
\end{align}
where $Tr$ denotes the functional trace:
\begin{align}
    Tr \left[ log \left( g^{00}\right) \right]=\int d^4 x \sqrt{g} \sum_i \phi_i^*(x) \phi_i(x) \log \left( g^{00}\right)=\int d^4 x \sqrt{g}  \log \left( g^{00}\right) \delta^{(4)}(0).
\end{align}
The value of $\delta^{(4)}(0)$ is ill defined, therefore it demands some regularization. For example in dimensional regularization $\delta^{(4)}(0)=0$, hence the measure in the functional integral is not sensitive to the factor $g^{00}$ \cite{Buchbinder:1992rb}.  But in the case of space-times with Killing horizons indefinite forms appear in the expression:
\begin{align}
    \delta^{(4)}(0) \log\left(g^{00}\right)\sim 0 \times \infty,
\end{align}
since $g^{00}$  vanish on the horizon. 

Further we will assume that the Euclidean functional integral is defined with a measure without a factor $g^{00}$, since in this case it is defined as the functional determinant \eqref{detpath}, and it is this approach that is widely discussed in the literature.

For the gaussian theory one-loop calculation is exact. There are several different methods to calculate the functional determinant. For example:

1) In terms of the $\zeta$-function (see e.g. \cite{  Vassilevich:2003xt,Hawking:1976ja}): 
\begin{align}
\log Z^E= \frac{1}{2}\left[\zeta'(0,\beta)+\zeta(0,\beta) \log(\mu^2)\right],\quad \text{where}\quad \zeta(s,\beta)= \sum_i \lambda_i^{-s}.
\nonumber
\end{align} 

2) In terms of the heat kernel:
\begin{align}
\log Z^E=\frac{1}{2} \int d^4 x\sqrt{g}  \int_{\epsilon^2}^\infty \frac{ds}{s} e^{-s m^2}  K(s,x,x) =\frac{1}{2}   \int_{\epsilon^2}^\infty \frac{ds}{s} e^{-s m^2} Tr\left[ K(s,x,x)\right],
\end{align}
where $\epsilon \sim 1/\mu$  and heat kernel $K(s,x,y)$ solves the following equation: 
\begin{gather}
    \partial_s K(s,x,y) +(-\Box_E+\xi R) K(s,x,y)=0 ,
\end{gather}
with the initial condition
\begin{align}
K(0,x,y)=\delta(x,y).
\end{align}
The heat kernel expansion in static space-times was widely discussed in the literature, see for example \cite{Birrell:1982ix,Fursaev:1995ef}.

Let us also stress that the heat kernel and zeta function methods are equivalent due to following relation: 
\begin{gather}
   e^{-s m^2} Tr\left[ K(s,x,x)\right] = \int d^4 x \sqrt{g }  \langle x| e^{-s (-\Box_E+m^2+\xi R)} |x\rangle =\\=
    \nonumber
  \sum_i \sum_j  \int dV \sqrt{g }  \langle x|\phi_i\rangle \langle \phi_i| e^{-s (-\Box_E+m^2+\xi R)} |\phi_j\rangle \langle \phi_j|x\rangle = \sum_i e^{- s\lambda_i}.
\end{gather}
Therefore, the effective action can be expressed in terms of the zeta function:
\begin{gather}
\log Z^E=\frac{1}{2} \sum_i  \int_{\epsilon^2}^\infty \frac{ds}{s} e^{-s m^2}  e^{- s\lambda_i}=-\frac{1}{2}\sum_i \log\left(\frac{\lambda_i}{1/\epsilon^2}\right) 
=\\= \frac{1}{2}\left[\zeta'(0,\beta)+\zeta(0,\beta) \log(\mu^2)\right].
\nonumber
\end{gather}
3)  In terms of the Feynman thermal propagator  \cite{Dowker:1975tf,Candelas:1975du}:
\begin{align}
\label{logZmassmethod}
\log Z=-\frac{1}{2}\int  d^4 x\sqrt{-g}\int_{\mu^2} ^{m^2} d\bar m^2  \ G (x,x),
\end{align}
where:  
\begin{align}
    G(x,y)=\sum_i \frac{\phi^*_i(x)\phi_i(y)}{\lambda_i}.
\end{align}
For more details on the Feynman thermal propagator in curved space-time, see Appendix \ref{appA}. This method is equivalent to the heat kernel and zeta function ones. It is easy to see due to the following integral representation of the thermal Feynman propagator:
\begin{gather}
G(x,y)=\big< x\big| \left(-\Box_E+m^2+\xi R\right)^{-1}\big|y\big> =  \int_0^\infty ds e^{-s m^2} \langle x| e^{-s (-\Box_E+\xi R)} |y\rangle 
\nonumber
=\\=
\int_0^\infty ds e^{-s m^2}  K(s,x,y).
\end{gather}

Hence, all standard calculation methods, such as the heat kernel, the zeta function and the integration of the thermal Feynman propagator at coincident points, are equivalent. Let us stress that all of these methods are used in the literature, аnd each of them is useful in different situations. In the next section we will use the last method \eqref{logZmassmethod}, to show that the Euclidean path integral and thermal partition function are equivalent only for the space-time without Killing horizon.

 \subsection{Thermal partition function}
The thermal partition function of the canonical ensemble is defined as:
\begin{align}
    Z^C=Tr(e^{-\beta :\hat{H}:}).
\end{align}
where $:\hat{H}:=\sum_i \omega_i \hat{a}^\dagger_i \hat{a}_i    $  is the usual normal ordered Hamiltonian defined with respect to the Killing vector $\partial_t$ and $\omega_k$ are the energies of the single-particle states. The trace is defined as the sum over all distinct (many-particle) states of the system:
\begin{align}
\label{tr}
Tr(e^{-\beta :\hat{H}:})= \prod_i\left(\sum_{n=0}^\infty e^{-\beta n \omega_i}\right)=\prod_i \left(1-e^{-\beta \omega_i}\right)^{-1}.
\end{align}
And the following relation is valid:
\begin{align}
     \langle :\hat{H}:\rangle =-\partial_\beta \log Z^C,
\end{align}
where
\begin{align}
    \hat{\rho}=e^{-\beta :\hat{H}:}/Z^C \quad \text{and}\quad  \langle \hat{O}\rangle =Tr(\hat{\rho} \hat{O}).
\end{align}

The energies of single-particle states can be found using Klein-Gordon equation (in Lorentzian signature) for the modes $\psi_k(x) =e^{-i \omega_k t} f_{\omega_k}(x)$:
\begin{align}
   \left(-\Box+m^2+\xi R\right) \psi_k(x)=0.
\end{align}
This equation can be rewritten as: 
\begin{align}
    g^{00}(\partial_t^2+H_s^2)e^{-i \omega_k t}f_{\omega_k}(x)= (-\omega_k^2+H_s^2)f_{\omega_k}(x)=0,
\end{align}
where $H_s$ is the quantum-mechanical single-particle Hamiltonian:
\begin{align}
\label{Hst}
    H^2_s=g_{00}\left(-\triangle_3+m^2+\xi R\right),
\end{align}
where $\triangle_3=g^{-\frac{1}{2}}\partial_i\left(g^{i j} g^{\frac{1}{2}}\partial_j\right) =\triangledown_i \triangledown^i+\frac{1}{2} (\triangledown^i \log |g_{00}|)\triangledown_i$, and its eigen-values determine single particle spectrum.

Now using the following factorizations: 
\begin{align}
\sinh\left( \frac{\beta\omega_n}{2} \right)=\frac{\beta\omega_n}{2} \prod_{k=1}^{\infty} \left(1+\frac{\beta^2\omega^2_n}{4 \pi^2 k^2}\right),
\end{align}
one can obtain the following identity:
\begin{align}
    \log\left(1-e^{-\beta \omega_n}\right)=-\frac{\beta\omega_n}{2}+\log(\beta\omega_n)+\sum_{k=1}^\infty \log \left(1+\frac{\beta^2\omega^2_n}{4 \pi^2 k^2}\right).
\end{align}
Then by using zeta function one can rewrite \eqref{tr} in the form:
\begin{align}
\label{cansum}
   \log Z^C= -\frac{1}{2}\sum_{n,k} \log\left(\frac{\frac{4\pi^2 k^2}{\beta^2}+\omega_n^2}{\mu^2}\right) + \beta \frac{1}{2}\sum_n \omega_n,   
\end{align}
where we have added normalization scale $\mu$ to make the logarithm dimensionless. The second term in the last equation is the zero point energy. This term is not very important, since zero point energy term is linear in inverse temperature and it does not affect the thermodynamic quantities. Hence, we can just remove this contribution. 

After the subtraction of the second term in \eqref{cansum}, the thermal partition function can be expressed in terms of the functional determinant \cite{Barbon:1994ej,Barbon:1995im} :\

\begin{align}
\label{detT}
   Z^C=  det^{- \frac{1}{2}}\left(\frac{-\partial_\tau^2+H_s^2}{\mu^2} \right)=  det^{-1/2}\left[\frac{g_{00}\left(-\Box_E+m^2+\xi R\right)}{\mu^2}\right],
\end{align}
For an optical manifold, $g_{00}=1$, therefore the thermal partition function is equal to the Euclidean path integral:
\begin{align}
   Z^C=  det^{-1/2}\left[\frac{\left(-\Box_E+m^2+\xi R\right)}{\mu^2}\right] = Z^E,
\end{align}
as was formulated in \cite{Allen:1986qi}. For more details of the formulation of statistical mechanics in ultastatic manifold see \cite{Dowker:1978md,Dowker:1988jw,Dowker:1989gp}.

To rewrite the partition function in terms of the Euclidean path integral we can use the basis of eigen-functions that solve the following equation:
\begin{align}
     g_{00}\left(-\Box_E+m^2+\xi R \right)\chi_i(x)=\rho_i \chi_i(x),
\end{align}
which are orthonormal with respect to the inner product:
\begin{align}
\label{innerT}
    \langle\chi_i| \chi_j\rangle = \int d^4x \sqrt{g}g^{00} \chi^*_1(x) \chi_2(x)=\delta_{ij}.
\end{align}
Then the corresponding action in terms of the inner product coincides with the standard covariant action: 
\begin{align}
    S[\chi]=\frac{1}{2}\langle\chi|  g_{00}\left(-\Box_E+m^2+\xi R \right) |\chi\rangle=\int d^4x \sqrt{g} \chi(x) \left(-\Box_E+m^2+\xi R \right) \chi(x). 
\end{align}
Since the inner product \eqref{innerT} does not coincide with the covariant one \eqref{innerq}, there is a difference in the path integral measure: 
\begin{align}
\label{detpartfunct}
Z^C=    \int D[\chi] e^{-S[\chi]}=  det^{-1/2}\left[\frac{g_{00}\left(-\Box_E+m^2+\xi R\right)}{\mu^2}\right],
\end{align}
where now:
\begin{align}
    D[\chi]=\prod_x d \chi(x) g^{\frac{1}{4}} (g^{00})^{\frac{1}{2}}.
\end{align}
This formal functional measure in general is non-covariant due to the explicit dependence on $g^{00}$. This point was discussed in \cite{Belgiorno:1995dt,deAlwis:1994ej,deAlwis:1995cia,Barbon:1995im}. 

Another representation of the thermal partition function \eqref{detT}  can be described in the optical approach \cite{deAlwis:1995cia}. By performing a conformal transformation from the static metric $g_{\mu \nu}$ to the related optical metric:
\begin{align}
    \bar{g}_{\mu \nu}=\Omega^2(x)g_{\mu \nu},
\end{align}
so that the interval and the field transform as:
\begin{align}
    d\bar{s}^2=\Omega^2(x) d s^2 \quad \text{and} \quad \bar{\varphi}(x)=\Omega^{-1}\varphi(x),
    \end{align}
where for the optical metric $\Omega^{-2}=g_{00}$. 
Under such a transformation the action changes to \cite{Birrell:1982ix}: 
\begin{align}
\label{Opticalop}
     \bar{S}_{\text{opt}}[\bar{\varphi}]=\frac{1}{2}\int d^4 x \sqrt{\bar{g}}
    \bar{\varphi}(x)\left[-\bar{\Box}_E +\frac{1}{6}\bar{R}+\Omega^{-2} m^2 +\Omega^{-2}\left(\xi-\frac{1}{6}\right)R \right]\bar{\varphi}(x)
\end{align} 
where the conformal transformation of the Ricci scalar is defined as follows:
\begin{align}
    \Omega^2 \bar{R}=R-6 \Omega^{-1} \triangledown_\mu\triangledown^\mu \Omega. 
\end{align}
Hence the operator in the action  \eqref{Opticalop} has the following form:
\begin{align}
    \left[-\bar{\Box}_E +\frac{1}{6}\bar{R}+\Omega^{-2} m^2 +\Omega^{-2}\left(\xi-\frac{1}{6}\right)R \right]=-\partial_\tau^2+\bar{H}_s^2,
\end{align}
where:
\begin{align}
\label{Hopt}
    \bar{H}_s^2=-\Omega^{-3} \triangledown_\mu\triangledown^\mu \Omega-2 \Omega^{-3} \triangledown_i \Omega \triangledown^i-\Omega^{-2} \triangledown_i  \triangledown^i+\Omega^{-2} m^2+\Omega^{-2}\xi R.
\end{align}
Now it is straightforward to check what the single-particle Hamiltonian for the theory in the optical metric \eqref{Hopt} can be expressed in terms of the single-particle Hamiltonian \eqref{Hst} of the original theory as follows \cite{Fursaev:1997th}: 
\begin{align}
    \bar{H}^2_s=\Omega^{-1} H^2_s \Omega.
\end{align}
From this relation it follows that the operators $H_s$ and $\bar{H}_s$ are equivalent and their eigen-values coincide.
As a result, the Euclidean path integral in the related optical manifold is equal to the partition function:
\begin{gather}
    Z^E_{\text{opt}} =\int d[\bar{\varphi}] e^{-\bar{S}_{\text{opt}}[\bar{\varphi}]}
    =\\=
        \nonumber
    det^{-1/2}\left[-\partial_t^2+\bar{H}_s^2\right]  = det^{-1/2}\left[-\partial_t^2+H_s^2\right]
    =\\=Tr(e^{-\beta :\hat{H}:})= Z^C.    
\end{gather}
In the case of compact manifolds, one can go further and prove that the Euclidean path integral in the original static metric is equal to the Euclidean path integral in the related optical manifold and therefore is equal to the partition function. Using the conformal transformation one can obtain:
\begin{align}
    Z^E=\int d[\varphi] e^{-S[\varphi]}=\int d[\bar{\varphi}] J(g,\bar{g},\beta) e^{-\bar{S}_{\text{opt}}[\bar{\varphi}]}.
\end{align}
The functional integration measure in the path integral is not invariant, and the transformation leads to the appearance of a functional Jacobian that does not depend on the value of the field. Therefore:
\begin{align}
    Z^E=J(g,\bar{g},\beta) Z^E_{\text{opt}}=J(g,\bar{g},\beta)Z^C.
\end{align}
For a compact and non-compact manifold without Killing horizon, the logarithm of the functional Jacobian is a linear function in inverse temperature, for more detail see \cite{Brown:1985ri,Dowker:1985mf}, and therefore they can be neglected in the renormalization of the free energy.  But in the case of space-times with Killing horizons, the Jacobian is ill defined, since $g_{00} \to 0$. Therefore the Euclidean path integral is equal to the thermal partition function for compact and non-compact manifolds without Killing horizon:
\begin{align}
    Z^E=Z^C.
\end{align}
Where under this equality we assume that they are equal to each other up to the exponent of a linear function of the inverse temperature. The main problem with this proof is that it is not obvious how to generalize it to space-times with Killing horizons. 

Below we will independently prove this statement and show why there is a difference between the Euclidean path integral and the thermal partition function for space-times with Killing horizons. 

Let us also stress that since the thermal partition function is equal to the Euclidean path integral which is generally covariant, then the thermal partition function is also generally covariant despite its explicitly non-covariant definition. But, it is not clear for us, whether the thermal partition function is generally covariant for non-compact manifolds with Killing horizons.

\section{Standard and new methods\label{sec3} }

In this section we discuss a method of computation of the Euclidean path integral in terms of the Feynman thermal propagator. We show that if the spectrum of the theory is discrete, i.e. for compact spaces with suitable boundary conditions, then the Euclidean path integral approach is equivalent to the thermal partition function one. The advantage of this method is that it uses the corresponding thermal Feynman propagator, and can be straightforwardly generalized to the case of non-compact spaces. However, the proof in the latter case is more complicated than in the case of compact spaces. Moreover, as we will see in the case of non-compact spaces with killing horizons the Euclidean path integral approach and thermal partition function one are not equivalent.

\subsection{Standard method: compact spaces }
As we explain in the Section \ref{sec2} one of the standard methods to calculate the Euclidean path integral is based on the integration of the thermal Feynman propagator \eqref{Gn} at coincident points:
\begin{align}
G(x,x)= \sum_i \frac{1}{2 \omega_i}    \phi_i(x)  \phi_i^*(x)\Big[1+ 2 n(\beta \omega_i)\Big],
\end{align}
over the mass and volume as in \eqref{logZmassmethod}. Then, the Euclidean path integral can be expressed as:
 \begin{align}
 \log Z^E = -\frac{1}{2}\beta\int_\infty^{m^2} d m^2  \int d^3 x \sqrt{g} \sum_i   \frac{1}{2 \omega_i}   \phi_i(x)  \phi_i ^*(x)\Big[1+ 2 n(\beta \omega_i)\Big].
 \end{align}
The first term in the integral on the right-hand side leads to the standard UV divergence due to zero-point fluctuations. It can either be subtracted or absorbed into the renormalization of the ground state energy after the regularization, since it is linear in the inverse temperature $\beta$.  Hence, we define:
 \begin{align}
 \log Z^E = -\beta\int_\infty^{m^2} d m^2  \int d^3 x \sqrt{g} \sum_i   \frac{1}{2 \omega_i}   \phi_i(x)  \phi_i ^*(x) n(\beta \omega_i).
 \end{align}

For the compact optical manifolds, $g^{00}=1$, the volume integral of the spatial part of the modes is equal to:
\begin{align}
 \int d^3 x \sqrt{g}    \phi_i(x)  \phi_i ^*(x) = 1 ,
 \end{align}
since it coincides with the orthogonality condition \eqref{ort}. Then from the equations of motion \eqref{eqmotion} it follows that the energy spectrum is defined as follows: $\omega_i^2=m^2+...$, therefore one can take the integral over the mass to obtain that:
\begin{align}
 \log Z^E = -\beta\int_\infty^{m^2} d m^2 \sum_i   \frac{1}{2 \omega_i}   n(\beta \omega_i)= - \sum_i  \log\left(1- e^{-\beta \omega_i}\right)=\log Z^C.
 \end{align}
For general compact static (but not optical) spaces let us use the following trick:
 \begin{align}
 \label{mm}
 I= \int d^3 x \sqrt{g}    \phi_i(x)  \phi_i ^*(x) = \lim_{m_1\to m_2} \frac{m_1^2-m_2^2}{m_1^2-m_2^2}  \int d^3 x \sqrt{g}    \phi_i(x,m_1)  \phi_i ^*(x,m_2),
 \end{align}
where $\phi_i(x,m_{1,2})$ solve the equations of motions with the corresponding masses: 
\begin{align}
(-g^{00}\omega_i^2(m_{1,2})-\triangle_3 +m_{1,2}^2+\xi R) \phi_i(x,m_{1,2})=0.
\end{align}
Here we indicate that each field $ \phi_i(x,m_{1,2})$ and the spectrum $\omega_i(m_{1,2})$ depend on the mass. 

Now one can rewrite the expression on the right hand side of \eqref{mm} as follows:
\begin{gather}
\label{twoterm}
    I= 
 \lim_{m_1\to m_2} \frac{ \omega_i^2(m_{1})- \omega_i^2(m_{2})}{m_1^2-m_2^2}  \int d^3 x \sqrt{g} g^{00}     \phi_i(x,m_1)  \phi_i ^*(x,m_2)
 +\\+
 \lim_{m_1\to m_2} \frac{1}{m_1^2-m_2^2}  \int d^3 x \sqrt{g}      \bigg[\triangle_3\phi_i(x,m_1)  \phi_i ^*(x,m_2) -  \phi_i(x,m_1)  \triangle_3 
 \phi_i ^*(x,m_2) \bigg] 
 \nonumber .
\end{gather}
As a result the Euclidean path integral takes the following form:
\begin{gather}
\label{logZ1}
 \log Z^E =  -  \beta\int d^3 x \sqrt{g} g^{00}\int_\infty^{m^2} d m^2       \sum_i   \phi_i(x)  \phi_i ^*(x)  \Big(\partial_{m^2} \omega_i\Big)    n(\beta \omega_i)
 + 
\\  
-
\beta \int d^3 x \sqrt{g} \int_\infty^{m^2}d m^2  \sum_i \frac{1}{2\omega_i}   \bigg[\triangle_3 \partial_{m^2}\phi_i(x)  \phi_i ^*(x) -  \partial_{m^2}\phi_i(x)  \triangle_3 
 \phi_i ^*(x) \bigg]n(\beta \omega_i) .
\nonumber 
\end{gather}
After taking the integral over the mass in the first term by parts, one gets:
\begin{gather}
\label{logZm1}
 \log Z^E =  -  \int d^3 x \sqrt{g} g^{00}      \sum_i   \phi_i(x)  \phi_i ^*(x)    \log\left(1- e^{-\beta \omega_i}\right)
 + \\+
 \nonumber
   \int d^3 x \sqrt{g} g^{00}      \sum_i  \int_\infty^{m^2}d m^2  \partial_{m^2} \left[\phi_i(x)  \phi_i ^*(x) \right]   \log\left(1- e^{-\beta \omega_i}\right)
-\\-
\beta \int d^3 x \sqrt{g} \int_\infty^{m^2}d m^2  \sum_i \frac{1}{2\omega_i}   \bigg[\triangle_3 \partial_{m^2}\phi_i(x)  \phi_i ^*(x) -  \partial_{m^2}\phi_i(x)  \triangle_3 
 \phi_i ^*(x) \bigg]n(\beta \omega_i) .
\nonumber 
\end{gather}
Let us label each term in the last expression as $\log Z^E_i$ where $i=1,2,3$ -- is the number of the line in \eqref{logZm1}. In the first term, one can take the volume integral by using the orthogonality relation \eqref{ort}. The second term vanishes if one uses again orthogonality relation:
\begin{align}
  \partial_{m^2}\int d^3 x \sqrt{g} g^{00}     \phi_i(x)  \phi_i ^*(x)= \partial_{m^2}1 =0.
\end{align}
The third term vanishes since its integrand is a total derivative and the field obeys suitable boundary conditions. Therefore, the Euclidean path integral is equal to the partition function for compact manifolds, because the spectrum of the theory is discrete: 
\begin{align}
 \log Z^E = \log Z^E_1 = -\beta\int_\infty^{m^2} d m^2   \sum_i    \frac{\partial \omega_i  }{\partial m^2}  n(\beta \omega_i)= - \sum_i  \log\left(1- e^{-\beta \omega_i}\right)=\log Z^C.
 \end{align}
Therefore, the fundamental statistical-mechanical relation \eqref{FSR} holds for compact manifolds. This statement has also been proven using the zeta function in \cite{Moretti:1997qn}. But it is not clear how to generalize the zeta function method to non-compact spaces, that is the reason why we consider other approaches. 

Now let us stress that for non-compact static spaces the spatial part of the mode has an incoming and a scattered wave, due to which the second and third terms in \eqref{twoterm}  do not vanish. But only the first term in \eqref{twoterm} leads to the equality between the Euclidean path integral and the thermal partition function. Below we will see that if the effective scattering potential for the spatial part of the mode has a finite range then the second and third terms in \eqref{twoterm} cancel each other.

Furthermore, if the space-time has Killing horizons where the metric is degenerate, then the spectrum of the theory does not depend on the mass, therefore there will be only the third term. We will discuss this case in the Section \ref{secHorizon}.

\subsection{Standard method: non-compact spaces }
Let us start with the definition of the partition function for non-compact spaces.   The trace in the definition of the thermal partition function is taken over all distinct (many-particle) states of the system \eqref{tr}, but it can be rewritten in terms of the trace over all single-particle excitation as follows:
\begin{align}
\log Tr(e^{-\beta :\hat{H}:})=- \sum_i \log \left(1-e^{-\beta \omega_i}\right) =-{Tr}_s \log\left(1-e^{-\beta \hat{H}_s}\right),
\end{align}
where we denote single-particle trace by the $Tr_s(\ \ )$ and $H_s$ is the quantum-mechanical single-particle Hamiltonian \eqref{Hst}. Then, using eigen-functions of the single-particle Hamiltonian \eqref{ort}, one can rewrite the trace as the volume integral:
\begin{align}
\label{partnoncomp}
\log Tr(e^{-\beta :\hat{H}:})=- \int d^3 x \sqrt{g} g^{00} \sumint_i      \phi_i(x)  \phi_i ^*(x) \log\left(1- e^{-\beta \omega_i}\right).
\end{align}
Here we assume that the spectrum of eigenvalues of the single-particle Hamiltonian can be discrete and continuous.  For compact spaces, one can take the volume integral to obtain exactly \eqref{tr}. But for non-compact spaces, one cannot take the volume integral in \eqref{partnoncomp}, since it will be proportional to $\delta(0)$. Therefore, one has to first take all the momentum integrals in order to express the thermal partition function through the volume integral of the free energy density: 
\begin{align}
    \log Z^C = - \int d^4 x \sqrt{g} F(\beta,x). 
\end{align}
Although, as we discussed in the Section \ref{sec2}, the partition function is not generally covariant due to the explicit dependence on $g_{00}$ in the corresponding functional determinant \eqref{detpartfunct}, it turned out to be covariant for compact spaces, since we showed that it is equivalent to the Euclidean path integral, which is generally covariant. Therefore, we assume that in non-compact spaces the free energy density $F(\beta,x)$ must be a generally covariantly function. But it does not have to be a local function. However it should depend on the geometric quantities ($g_{\mu \nu},R,R_{\mu\nu},...)$, covariant derivatives and a four-temperature $\beta_\mu$ \footnote{In curved space-times or curve-linear coordinates the temperature measured by a co-moving thermometer is equal to \cite{Tolman:1930ona, Tolman:1930zza}: 
\begin{align}
    T_{local}=  \frac{T}{\sqrt{g_{00}}}.
\end{align}
And the most natural way to describe thermodynamic quantities is in term of the so called four-temperature:
\begin{align}
    \beta_\mu =\beta \xi_\mu 
\end{align}
where $\xi_\mu$ is the time-like Killing vector. Therefore the combination $\beta_\mu\beta^\mu=g_{00}\beta^2$ is a generally covariant object. } ?

In non-compact spaces in general the partition function may diverge due to the infinite space-time volume or, as we will see in the next section, due to the presence of Killing horizon, even if the volume of the space-time is finite.

As we have shown in the previous subsection in the case of non-compact manifolds the Euclidean path integral contains three terms:
\begin{gather}
\label{logZm}
 \log Z^E =  -  \int d^3 x \sqrt{g} g^{00}      \sumint_i   \phi_i(x)  \phi_i ^*(x)    \log\left(1- e^{-\beta \omega_i}\right)
 + \\+
 \nonumber
   \int d^3 x \sqrt{g} g^{00}      \sumint_i  \int_\infty^{m^2}d m^2  \partial_{m^2} \left[\phi_i(x)  \phi_i ^*(x) \right]   \log\left(1- e^{-\beta \omega_i}\right)
-\\-
\beta \int d^3 x \sqrt{g} \int_\infty^{m^2}d m^2  \sumint_i \frac{1}{2\omega_i}   \left[\triangle_3 \partial_{m^2}\phi_i(x)  \phi_i ^*(x) -  \partial_{m^2}\phi_i(x)  \triangle_3 
 \phi_i ^*(x) \right]n(\beta \omega_i) .
\nonumber 
\end{gather}
The first term $\log Z^E_1$ is equal to the standard definition of the partition function \eqref{partnoncomp}. 

The second term may seem to be zero, since if one takes the integral over the spatial coordinates, one obtains $\partial_{m^2}\delta_i(0)$, but the derivative of the delta function at the origin is ill-defined. Furthermore, if the energy spectrum of the theory does not depend on the mass $\partial_{m^2} \omega_i=0$, then the sum of the first and second terms vanishes, as follows from \eqref{logZ1}. Therefore the second term demands a careful attention. In the case when the spatial part of the mode is a solution of the Schrodinger equation with a scattering potential, the spatial part of the mode depends on the mass. Because of that the third term also does not vanish. 

For non-compact optical manifold, $g_{00}=1$, the equality between the Euclidean path integral and the thermal partition function is obvious, since the spatial part of the mode does not depend on the mass, and the second and third terms in \eqref{logZm} vanish. 

To illustrate contributions of different terms in \eqref{logZm} let us consider a model  metric of the following form:
\begin{align}
\label{metric}
    ds^2=(1+f(x))(-dt^2+dx^2)+d\vec{z}^2.
\end{align}
where $\lim_{x\to \pm \infty}f(x)=0$. This metric is not optical, but as we will see, the second and third terms cancel each other if the effective potential for the spatial parts for the modes has a finite range.

The equation of motion for the free massive scalar field is:
\begin{align}
(g^{00}\partial_t^2-\triangle_3  +m^2+\xi R) \varphi=0.
\end{align}
Then the corresponding quantum field operator has the following form:
\begin{align}
\label{scatphi}
    \hat{\varphi} = \int_0^\infty \frac{dp}{\sqrt{2 \pi}} \int \frac{d^2k}{(\sqrt{2 \pi})^2} \frac{e^{-i \omega t}}{\sqrt{2 \omega}} e^{i \vec{k}\cdot \vec{z}}\left[\overset{\rightarrow}{\phi}_p(x) \hat{a}_{p,\vec{k}}+\overset{\leftarrow}{\phi}_{p}(x) \hat{b}_{p,\vec{k}}\right]+h.c.,
\end{align}
where $\omega=\sqrt{m^2+p^2+\vec{k}^2}$ and  $\overset{\rightarrow}{\phi}_{p}(x),\overset{\leftarrow}{\phi}_{p}(x)$ are the scattering eigen-functions of the effective Schrodinger equation:
\begin{gather}
    \left[-\partial_x^2  +\mathbf{V} \right]\overset{\leftarrow}{\overset{\rightarrow}{\phi}}_{p}(x) =p^2  \overset{\leftarrow}{\overset{\rightarrow}{\phi}}_{p}(x),
\end{gather}

with the effective potential $ \mathbf{V}=f(x)(k^2+m^2)+(1+f(x))\xi R$, which vanishes at spatial infinity.

As it is known, solutions for such a potential contain the usual asymptotic plane waves with reflected and transmitted components. The asymptotics of the right moving waves are:
\begin{align}
\label{phiright}
   \overset{\rightarrow}{ \phi}_{ p}(x) \approx  \theta(-x)\left(e^{i p x}+\overset{\rightarrow}{R}_p e^{-i p x}\right)+\theta(x) \overset{\rightarrow}{T}_p e^{i p x}, 
\end{align}
while of the left moving waves are:
\begin{align}
\label{phileft}
    \overset{\leftarrow}{\phi}_{p}(x) \approx  \theta(-x) \overset{\leftarrow}{T}_{p} e^{-i p x} +\theta(x) \left(e^{-i p x}+\overset{\leftarrow}{R}_{p} e^{i p x}\right),
\end{align}
where the reflection and transmission amplitudes obey the following relations: 
\begin{align}
\label{RT}
     \overset{\rightarrow}{T}_p= \overset{\leftarrow}{T}_p,
     \quad
     \overset{\rightarrow}{R}_p  \overset{\rightarrow}{T^*}_p=- \overset{\leftarrow}{R^*}_p  \ \overset{\leftarrow}{T}_p
     \quad \text{and} \quad 
     \big| \overset{\leftarrow}{\overset{\rightarrow}{R}}_p \big|^2 +  \big| \overset{\leftarrow}{\overset{\rightarrow}{T}}_p \big|^2=1.
    \end{align}
    
The reflection and transition amplitudes must somehow depend on the energy of the scattering waves and the geometric properties (for example the volume) of the space. If the potential substantially changes in an interval $[-a,a]$ and decreases to a constant, then for $|x|>a$  the spatial parts of the mode take the asymptotic form \eqref{phiright} and \eqref{phileft}. Let us now show that the second and third terms in \eqref{logZm} depend only on the region where the potential is not zero, i.e., both of them vanish in the region where the modes have such asymptotic forms as \eqref{phiright} and \eqref{phileft}.  

For the second term one can show that for the interval $(a,+\infty)$:
\begin{gather}
  \int_a^\infty d x      \int_{-\infty}^\infty d p   \partial_{m^2} \left[\overset{\rightarrow}{ \phi}_p(x)  \overset{\rightarrow}{ \phi}_p^*(x)+\overset{\leftarrow}{ \phi}_p(x)  \overset{\leftarrow}{ \phi}_p^*(x) \right]
  \approx\\ \approx \nonumber \int_a^\infty d x      \int_{-\infty}^\infty d p    \partial_{m^2}  \left[1+R_p e^{-2 i p x} \right]
    =  
     \int_a^\infty d x      \int_{-\infty}^\infty d p    \partial_{m^2} R_p e^{-2 i p x}.
\end{gather}
The last term is finite, therefore one can safely change the integration over the momentum and spatial coordinate. Hence:
\begin{align}
    \nonumber
     \int_a^\infty d x      \int_{-\infty}^\infty d p   \partial_{m^2} R_p e^{-2 i p x} \sim    \int_{-\infty}^\infty d p    \partial_{m^2} R_p \delta(p)= \partial_{m^2} R_{p=0}=0,
\end{align}
where in the last step we have used that the reflection amplitude obeys $R_{p=0}=-1$. The same is true for the interval $(-\infty,-a)$.

For the third term in \eqref{logZm}, it is straightforward to see that in the region where the spatial part of the mode takes asymptotic form  \eqref{phiright} and \eqref{phileft}, the integrand of $\log Z^E_3$  vanishes:
\begin{align}
   \bigg[\Big(\triangle_3 \partial_{m^2} \overset{\rightarrow}{ \phi}_p(x)\Big)  \overset{\rightarrow}{ \phi}_p^*(x) -  \Big(\partial_{m^2}\overset{\rightarrow}{ \phi}_p(x)\Big)  \triangle_3 
 \overset{\rightarrow}{ \phi}_p^*(x) \bigg] =0.
\end{align}
Therefore, the second and third terms receive contributions only from the region in which the potential is not zero, and they are finite. This means that these terms cannot be proportional to the the volume of the whole space-time, as it is the case for the thermal partition function. Hence, one can safely change the integration over momentum and spatial coordinates in both terms. Therefore, using \eqref{RT}, one can show that the third term can be written in the form: 
\begin{gather}
    \label{phas}
\log Z_3^E=-
\beta\int_\infty^{m^2}d m^2 \int d^3 x \sqrt{g}   \int_{-\infty}^\infty \frac{d^2 k}{(2\pi)^2} \int_0^\infty \frac{d p}{2\pi}  \frac{1 }{2 \omega}   n(\beta \omega)
\times \\ \times 
\Bigg( 
\bigg[\Big(\triangle_3 \partial_{m^2} \overset{\rightarrow}{ \phi}_p(x)\Big)  \overset{\rightarrow}{ \phi}_p^*(x) -  \Big(\partial_{m^2}\overset{\rightarrow}{ \phi}_p(x)\Big)  \triangle_3 
 \overset{\rightarrow}{ \phi}_p^*(x) \bigg]+ \bigg[
 \overset{\rightarrow}{ \phi}\to \overset{\leftarrow}{ \phi}\bigg]\Bigg) 
 \nonumber
 = \\ =
\nonumber
\beta A \int_\infty^{m^2}d m^2 \int_{-\infty}^\infty \frac{d^2 k}{(2\pi)^2} \int_0^\infty \frac{d p}{2\pi}  \frac{2 p}{2 \omega}   n(\beta \omega) \partial_{m^2} \left[  \theta_{\overset{\rightarrow}{R}_p}+ \theta_{\overset{\leftarrow}{R}_p}\right],
\end{gather}
where $A$ is the volume of the transverse directions in \eqref{metric} and $\theta_{\overset{\rightarrow}{R}_p}, \theta_{\overset{\leftarrow}{R}_p}$ are the phases of the reflection amplitudes at spatial infinities. Therefore the third term depends only on the scattering phase.

Now we can use the Friedel formula that connects the integrated density of states and the energy derivative of scattering phaseshifts (see e.g. \cite{Guo:2022row}) to obtain that:
\begin{gather}
    \int_{-\infty}^\infty dx \left( \left[\overset{\rightarrow}{ \phi}_p(x)  \overset{\rightarrow}{ \phi}_p^*(x)+\overset{\leftarrow}{ \phi}_p(x)  \overset{\leftarrow}{ \phi}_p^*(x) \right]
    -
    \left[\overset{\rightarrow}{ \phi_0}_p(x)  \overset{\rightarrow}{ \phi_0}_p^*(x)+\overset{\leftarrow}{ \phi_0}_p(x)  \overset{\leftarrow}{ \phi_0}_p^*(x) \right] \right) 
    =\\= \nonumber 
  \frac{d}{dp} \left[\theta_{\overset{\rightarrow}{R}_p}+ \theta_{\overset{\leftarrow}{R}_p}\right],
\end{gather}
where $\overset{\rightarrow}{ \phi_0}_p(x)=e^{i p x} $ and $\overset{\leftarrow}{ \phi_0}_p(x)=e^{-i p x} $ are the modes for the case of the absence of the scattering potential, i.e. for the flat space-time. 

Taking the integral over $p$ by part in \eqref{phas} and using the Friedel formula one can show that the second and third terms in \eqref{logZm} cancel each other: 
\begin{align}
    \log Z_2^E+\log Z_3^E=0.
\end{align}
Therefore, the Euclidean path integral is equal to the partition function for non-compact manifolds without Killing horizons. 

Let us note that our reasoning in this section works because we can change the integrals over momenta and coordinates, since these contributions are finite. But in the case of space-times with Killing horizons the Euclidean path integral is fully determined by the third term in \eqref{logZm}, and in that case these contribution diverge, which generally speaking, means that we cannot change the order of integrations. We will discuss that in the next section.

\subsection{New method\label{secNewmethod}} 

In this section, we propose a new method of computation of the Euclidean path integral that gives the correct answer for compact and non-compact spaces. This method gives the correct equality between the Euclidean path integral and the thermal partition function even for non-compact manifolds with Killing horizons. Therefore, this method seems to be suitable for computation of thermodynamic quantities in terms of the Euclidean path integral for a generic case. 

Let us introduce a new mass term, $M$, into the action \eqref{S}, which depends on $g^{00}$:
\begin{align}
    S_M[\varphi]=-\frac12\int dV \sqrt{-g}\Big[(\partial_\mu\varphi)^2+ g^{00}M^2\varphi^2+m^2\varphi^2+ \xi R \varphi^2 \Big].
\end{align}
This action is, generally speaking, not covariant, although in the limit $M\to 0$ the general covariance is restored and the following relation is valid: 
\begin{align}
    Z^E=\int d[\varphi]e^{-S} =\lim_{M\to 0} \int d[\varphi]e^{-S_M}=\lim_{M\to 0} Z^E_M.
\end{align}
This allows us to express the new Euclidean path integral via the integral with respect to $M^2$ rather than $m^2$, of the Feynman thermal propagator in the coincidence limit:
\begin{align}
\log Z_\beta^M=-\frac{\beta}{2}\int d^3 x\sqrt{g} g^{00}\int_\infty^{M^2} d M^2  \ G_M (x,x),
\label{Seff}
\end{align}
where: 
\begin{align}
G_M(x,x)= \sumint_i\frac{1}{2 \sqrt{\omega_i^2+M^2}}    \phi_i(x)  \phi_i^*(x)\left[ 1+ 2 n\left(\beta \sqrt{\omega_i^2+M^2}\right)\right].
\end{align}
Compare this expression to \eqref{logZmassmethod}. The spatial part of the mode $\phi_i(x)$ does not depend on $M$, since the new mass term $g^{00}M^2\varphi^2$ in the action simply shifts the energy spectrum, because it is the same as the temporal contribution $g^{00}(\partial_t \varphi)^2$. (Let us also stress, that modes obey the standard orthogonality and completeness relations.) Therefore the Euclidean path integral can be expressed as follows:
 \begin{align}
\log Z^E_M = -\beta \int d^3 x \sqrt{g} g^{00}\frac{1}{2}\int_\infty^{M^2} d M^2   \sumint_i \frac{1}{2 \sqrt{\omega_i^2+M^2}}   \phi_i(x)  \phi_i^*(x)\Bigg[1+ 2 n\bigg(\beta \sqrt{\omega_i^2+M^2}\bigg)\bigg].
 \end{align}
The first term under the integral on the right-hand side of the last equation leads to the standard UV divergence due to the zero-point fluctuations. This term is linear in inverse temperature. Hence, it can be absorbed into the renormalization of the ground state energy. As a result:
\begin{align}
\log Z^E=\lim_{M\to 0} \log Z^E_M = -\beta\int d^3 x \sqrt{g} g^{00}\int_\infty^{0} d M^2  \sumint_i \frac{1}{2 \sqrt{\omega_i^2+M^2}}  \phi_i(x)  \phi_i^*(x) \ n\left(\beta \sqrt{\omega_i^2+M^2}\right).
 \end{align}
 
Now one can take the integral over $M$ since the spatial part of the mode $\phi_i(x)$ does not depend on $M$:
\begin{align}
    \int_\infty^{0} d M^2   \frac{1}{\sqrt{\omega_i^2+M^2}}  n\left(\beta \sqrt{\omega_i^2+M^2}\right) = \frac{2}{\beta}\log\left(1- e^{-\beta \omega_i}\right).
\end{align}
Therefore:
\begin{align}
\log Z^E = - \int d^3 x \sqrt{g} g^{00} \sumint_i \phi_i(x)  \phi_i^*(x) \log\left(1- e^{-\beta \omega_i}\right)=\log Z^C.
 \end{align}
 Thus, the new method does indeed work.

\section{Spacе-times with horizons\label{secHorizon}} 
Due to Hawking type of radiation \cite{Hawking:1975vcx} space-times with horizons are usually endowed with a natural (canonical) temperature, which depends on the geometry of the space-time. Moreover, there is a Bekenstein-Hawking entropy \cite{Bekenstein:1972tm, Bekenstein:1973ur,Hawking:1975vcx,Gibbons:1977mu}, which is proportional to the area of the horizon and depends on the geometric properties of the whole space-time. In space-times with Killing horizons, one can also consider a thermal gas with the planckian density matrix with an arbitrary temperature, different from the canonical one. But if the temperature is different from the canonical one, then correlation functions do not posses Hadamard properties on Killing horizons  and the back-reaction on the background geometry is strong (see, for example: \cite{Candelas:1980zt, Fulling:1977zs,Akhmedov:2019esv, Akhmedov:2020ryq,Akhmedov:2020qxd, Anempodistov:2020oki, Akhmedov:2021cwh, Bazarov:2021rrb,Akhmedov:2022qpu}). Furthermore, the thermalization process in curved space-times, in general, is far from being well understood \cite{Akhmedov:2021rhq}.  
Let us also stress that if the temperature is less than the Hawking one, then the theory becomes unstable due to the presence of a tachyon excitation \cite{Diakonov:2023hzg}.

In any cases, in order to understand the general thermal properties in curved space-times thermal states with generic temperatures should be considered. For example, to find thermodynamic quantities, one needs to know derivatives of the thermal partition function with respect to the temperature. Therefore it is necessary to know the value of the free energy not only for the canonical temperature.

The other difference between non-compact space-times with Killing horizons and without them is that the spectrum of the theory in the former case does not depend on the mass. Therefore, the Euclidean path integral is defined only by the third term in \eqref{logZm}, which does not give the equivalence between the Euclidean path integral and thermal partition function approaches. Nevertheless here we show that the new method proposed in the previous section gives the correct equality.  

But before we begin to discuss the relation between the Euclidean path integral and thermal partition function approaches let us start with the definition of the energy. As shown in \cite{Fursaev:1997th}, it is necessary to distinguish different definitions of energy in space-times with Killing horizons. We will show that the difference between the energy defined by the stress-energy tensor ($E$) and the canonical Hamiltonian ($H_c$) is a boundary term ($Q_\xi$), which depends on the coupling constant $\xi$ in \eqref{S}, for the related discussion see also \cite{Iyer:1995kg,Frolov:1998vs}. Furthermore, we will show that the operator of the canonical Hamiltonian is different from the standard one $\hat{H}=\sum_i \omega_i \hat{a}^\dagger_i \hat{a}_i$ by another boundary term ($Q_H$). All these observations are important to establish the proper relation between the Euclidean path integral and the thermal partition function in space-times with Killing horizons.

The variation of the action of the massive non-minimally coupled scalar field theory \eqref{S} with respect to the field and metric gives equation of motion:
\begin{align}
\label{Box}
    (-\Box +m^2+\xi R) \varphi(x)=0 ,
\end{align}
and the stress energy tensor is defined as:
\begin{gather}
    T_{\mu \nu}= \partial_\mu \varphi(x) \partial_\nu \varphi(x) - \frac{1}{2} g_{\mu \nu}\left(\partial_\rho \varphi(x) \partial^\rho \varphi(x)+m^2 \varphi^2(x)   \right)
   \nonumber +\\+
   \xi\left[ \left(R_{\mu \nu}-\frac{1}{2}g_{\mu \nu} R\right)\varphi^2(x)+g_{\mu \nu} \Box \varphi^2(x)-\triangledown_\mu\triangledown_\nu \varphi^2(x)\right].
\end{gather}
Then the energy operator of the system can be defined in terms of the stress-energy tensor:
\begin{align}
    E=-\int d^3 x \sqrt{g}  T_{0}^0.
\end{align}
At the same time the canonical Hamiltonian is defined as follows: 
\begin{align}
   H_c=\int d^3x\left(\partial_0 \varphi(x) \pi (x)-L\right)=\int d^3x \sqrt{g} \mathbf{H},
\end{align}
where $L$ is the Lagrangian and the canonical Hamiltonian density is:
\begin{align}
    \mathbf{H}=\frac{1}{2}\left[-g^{00}\partial_0\varphi(x)\partial_0\varphi(x)+g^{i j}\partial_i\varphi(x)\partial_j\varphi(x)+(m^2+\xi R)\varphi^2(x)\right].
\end{align}
As one can see, even on the classical level there is a difference between these two definition of the energy density for non zero $\xi$: 
\begin{align}
    -T_0^0=\mathbf{H}-\xi\left(R_0^0\varphi^2(x)+\triangledown_i \triangledown^i\varphi(x)\right).
\end{align}
For a static space-time one can show that the second term is the total derivative: 
\begin{align}
    -T_0^0=\mathbf{H}-\xi \frac{1}{\sqrt{g}}\partial_i\left[ \sqrt{g} g^{i j}\left( \partial_j \phi^2(x)-\phi^2(x)w_i\right)\right],
\end{align}
where we use that for static space-times $R_0^0=-\triangledown^i w_i$ and  $w_i=\frac{1}{2}\partial_i \log|g_{00}|$. Hence, the difference between the energy and the Hamiltonian is: 
\begin{align}
\label{Ehxi}
    E=H_c -\xi \int d A^i \sqrt{|g_{00}|} \left( \partial_i (\phi^2(x))-\phi^2(x)w_i\right).
\end{align}
If the field obeys suitable boundary conditions then the energy is equal to the Hamiltonian. But if the boundary is a Killing horizon then the last term in \eqref{Ehxi} does not vanish as we will see.

Furthermore using the standard canonical quantization procedure, one can represent the field operator in static space-times as: 
\begin{align}
    \varphi(x)=\sumint_i e^{-i \omega_i t}\phi_i(x) \hat{a}_i + h.c.,
\end{align}
where $e^{-i \omega_i t} \phi_i(x)$ solves Klein-Gordon equation \eqref{Box}. Then the Hamiltonian operator is defined as follows: 
\begin{align}
\label{Ham}
    \hat{H}=\sumint_i \omega_i \hat{a}^\dagger_i \hat{a}_i,
\end{align}
and the following Heisenberg equation are valid: 
\begin{align}
\label{comrelation}
     i\partial_t\hat{\varphi(x)} = [\hat{\varphi}(x),\hat{H}]\quad \text{,}\quad i\partial_t\hat{\pi}(x) = [\hat{\pi}(x),\hat{H}]\quad \text{and }\quad [\hat{\varphi}(x),\hat{\pi}(y)]=\delta^{(3)}(x-y).
\end{align}
Now let us look at the standard definition of the canonical Hamiltonian ( for more details see the discussion around eq.\eqref{boundHam}): 
\begin{gather}
\label{canHam}
\hat{H}_c
=
\sumint_i \ \omega_i \ \hat{a}^\dagger_{\omega,k} \hat{a}_{\omega,k}+\frac{1}{4}\int d^3 x \partial_i[g^{i j}\sqrt{g}\partial_j \hat{\varphi}^2(x)].
\end{gather}
For space-times without Killing horizons the last term vanishes, therefore the canonical Hamiltonian operator has the correct form. But if the space-time has a Killing horizon, then the canonical Hamiltonian has a boundary term, whose expectation value is not zero. One should define the Hamiltonian operator as \eqref{Ham}, rather than \eqref{canHam} to establish the proper equations of motion following from \eqref{comrelation}. 

Therefore, we have the following relations:
\begin{align}
\label{BQ}
    \hat{H}_c=\hat{H}+\hat{B}_H \quad \text{and }\quad \hat{E}= \hat{H}_c+\hat{Q_\xi},
\end{align}
where:
\begin{align}
\label{Q}
    \hat{B}_H=\frac{1}{4}\int d A^i \sqrt{|g_{00}|} \partial_i \hat{\phi}^2(x) \quad \text{and }\quad  \hat{Q}_\xi= -\xi \int d A^i \sqrt{|g_{00}|} \left[ \partial_i \hat{\phi}^2(x)-\hat{\phi}^2(x)w_i\right].
\end{align}
Let us stress that $Q_\xi$ is the Noether charge, since there is a conserved current \cite{Iyer:1995kg,Fursaev:1997th,Frolov:1998vs}: 
\begin{align}
    J_\mu=-\xi\left(R_{\mu \nu}\varphi^2(x)+g_{\mu \nu}\Box \varphi^2(x)-\triangledown_\mu\triangledown_\nu\varphi^2(x)\right) \zeta^\nu,
\end{align}
where $\zeta^\nu$ is the Killing vector.  At the same time $B_H$ is not a charge of some conserved current.

\subsection{Rindler coordinates}
Now let us find explicitly the expectation values of the three energy operators $ \hat{H}_c,\hat{H},\hat{E}$, which we have introduced in \eqref{BQ}, and the expectation value of both boundary terms $\hat{B}_H,\hat{Q}_\xi$ in space-times with Killing horizons. For simplicity, let us consider only the Rindler coordinates, which approximate the region near the horizon of any space-time:
\begin{align}
ds^2 = r^2d t^2 -d r^2 - d \vec{z}^2,
\end{align}
 where we set the acceleration to one. 
 
 The field operator is defined as follows:
\begin{align}
\label{modeRindler}
    \hat{\varphi}(x)=\int \frac{d^2 k}{(2\pi)}\int \frac{d\omega}{\sqrt{\pi}}\sqrt{\frac{\omega \sinh(\pi \omega)}{\pi}} e^{-i \omega t+i \vec{k}\vec{x}}K_{i \omega}\left(\sqrt{m^2+k^2} r\right) \hat{a}_\omega+h.c.,
\end{align} 
and obeys the standard canonical commutation relations. 

Then for the massless case, the expectation value of the Hamiltonian can be expressed as: 
\begin{gather}
   \langle \hat{H}\rangle =
  \frac{1}{2}\int d^3 x \sqrt{g} g^{00} \lim_{x_0\to y_0}(\partial^2_{y_0}-\partial_{x_0}\partial_{y_0})  G(x,y) ,
\end{gather}
where the thermal Feynman propagator with inverse temperature $\beta$ is (for more details see Appendix \ref{appA}): 
\begin{gather}
\label{green0RIn}
    G(x_1,x_2)
    = \\=
    \nonumber
    \frac{1}{\beta}\sum_{\omega_n}\int \frac{d^2 \vec{k}}{(2 \pi)^2} \int_0^{\infty} \frac{ d \omega}{\pi^2} \frac{2\omega\sinh\pi \omega}{w_n^2+\omega^2}e^{-iw_n (\tau_2-\tau_1)}e^{i\vec{k}(\vec{z}_2-\vec{z}_1)}K_{i \omega}\big(\sqrt{m^2+k^2} r_1\big)K_{i\omega}\big(\sqrt{m^2+k^2} r_2\big),
\end{gather}
where $w_n=\frac{2\pi n}{\beta}$ are Matsubara frequencies. It should be kept in mind that there is a certain temperature – so called canonical (Unruh) temperature $T^{-1}_c=\beta_c=2\pi$, for which the poragator and loop corrections respect Poincaré symmetry \cite{Higuchi:2020swc,Akhmedov:2022uug}.

Hence, one obtains that the regularized energy is:
\begin{align}
\label{expH}
   \langle \hat{H}\rangle=  \int d^3 x \sqrt{g}  \frac{\pi^2}{30} \bigg[ \frac{1}{(\beta_\mu\beta^\mu)^2}-\frac{1}{((\beta_c)_\mu(\beta_c)^\mu)^2}\bigg].
\end{align}
where $\beta^\mu_c=\beta_c \xi^\mu$ is the four-vector of inverse canonical temperature, $\xi^\mu=(1,0,0,0)$ is the time-like Killing vector and $\beta^\mu=\beta \xi^\mu$ is the four-vector of inverse temperature of the state under consideration. Then, for the canonical temperature the expectation value of the energy vanishes. This is consistent with the fact that for the Poincaré invariant state (Minkowski vacuum), the expectation value of the energy should be zero. 

If we introduce a cutoff $r=\epsilon$ for the integration over the proper distance from the horizon in \eqref{expH}, then:

\begin{align}
\label{HV}
   \langle \hat{H}\rangle  = A\frac{1}{2\epsilon^2}\frac{\pi^2}{30}\bigg[ \frac{1}{\beta^4}-\frac{1}{\beta_c^4}\bigg].
\end{align}
where $A$ is the area of the horizon. This value is divergent in the limit $\epsilon\to 0$. This divergence has a physical meaning. For example, in the entire Minkowski space the expectation value of the Hamiltonian is $\langle \hat{H}\rangle=V \frac{\pi^2}{30} \frac{1}{\beta^4}$, and is also divergent due to an infinite volume of space. But the energy density is finite. Hence we prefer to understand the value on the energy as an integral over the volume of the density in \eqref{expH} rather than divergent value of the whole energy \eqref{HV}. It should be kept in mind that the energy density in space-times with Killing horizons diverges due to the fact that the observer who is fixed near the horizon will see that the local temperature is $T_{local}=\frac{T}{\sqrt{g_{00}}}$, which diverges with the decreases of the distance to the horizon, due to the infinite blue shift. Thus, the presence of the divergence under consideration is quite natural to space-times with Killing horizons. 

For the non-minimally coupled massless scalar field the expectation value of the stress energy tensor is:
\begin{gather}
\langle \hat{T}^{\mu}_\nu\rangle
=\\=
\nonumber
\frac{\pi^2}{90}\frac{1}{r^4}\Bigg( \bigg[ \frac{1}{\beta^4}-\frac{1}{\beta_c^4}\bigg] diag(-3,1,1,1)
+ 20 (6 \xi -1)\frac{1}{\beta_c^2}\bigg[ \frac{1}{\beta^2}-\frac{1}{\beta_c^2}\bigg]diag(3/2,-1/2,1,1) \bigg).
\end{gather}
This result can be obtained directly by using $\zeta$-function approach \cite{Moretti:1997qn} or point splitting \cite{Frolov:1987dz}. 

Then the energy of the system following from this expectation value is:
\begin{gather}
\langle \hat{E}\rangle =-\int d^3 x \sqrt{g} \langle T_{0}^0 \rangle
\nonumber
=\\=
\nonumber
 \int d^3 X \sqrt{g} \Bigg( \frac{\pi^2}{30} \bigg[ \frac{1}{(\beta_\mu\beta^\mu)^2}-\frac{1}{((\beta_c)_\mu(\beta_c)^\mu)^2}\bigg]
- (6 \xi -1) \frac{\pi^2}{3} \frac{1}{((\beta_c)_\mu(\beta_c)^\mu)}  \bigg[ \frac{1}{\beta_\mu\beta^\mu}-\frac{1}{(\beta_c)_\mu(\beta_c)^\mu}\bigg] \Bigg)
= \\=
\label{Exi}
A \frac{1}{2 \epsilon^2}\frac{\pi^2}{30} \bigg[ \frac{1}{\beta^4}-\frac{1}{\beta_c^4}\bigg]  
- (6 \xi -1) A \frac{1}{2 \epsilon^2}\frac{\pi^2}{3}
\frac{1}{\beta_c^2}\bigg[ \frac{1}{\beta^2}-\frac{1}{\beta_c^2}\bigg] .
\end{gather}
Then using the relation:
\begin{align}
    \langle\hat{E}\rangle= \langle\hat{H}\rangle+\langle\hat{Q}_\xi\rangle+\langle\hat{B}_H\rangle,
\end{align}
one can obtain the expectation values of the charge and boundary terms:

\begin{gather}
\label{QRin}
\langle\hat{Q}_\xi\rangle
=\\=
\nonumber
 -\xi\int d^3 X \sqrt{g}  2\pi^2  \frac{1}{((\beta_c)_\mu(\beta_c)^\mu)}  \bigg[ \frac{1}{\beta_\mu\beta^\mu}-\frac{1}{(\beta_c)_\mu(\beta_c)^\mu}\bigg] 
= - \xi A \frac{1}{2 \epsilon^2}2\pi^2
\frac{1}{\beta_c^2}\bigg[ \frac{1}{\beta^2}-\frac{1}{\beta_c^2}\bigg] 
\end{gather}
and 
\begin{gather}
\label{BRin}
\langle\hat{B}_H\rangle 
=\\=
\nonumber
 \int d^3 X \sqrt{g}  \frac{\pi^2}{3}  \frac{1}{((\beta_c)_\mu(\beta_c)^\mu)}  \bigg[ \frac{1}{\beta_\mu\beta^\mu}-\frac{1}{(\beta_c)_\mu(\beta_c)^\mu}\bigg] 
=  A \frac{1}{2 \epsilon^2}
\frac{1}{\beta_c^2}\bigg[ \frac{1}{\beta^2}-\frac{1}{\beta_c^2}\bigg].
\end{gather}
This result can be obtained directly from the definition \eqref{Q} of these operators and from the expectation value of the regularized Feynman thermal propagator at coincident points \cite{Diakonov:2023hzg}:
\begin{align}
    \langle \hat{\phi}^2(x) \rangle= \frac{1}{12 r^2}\left(\frac{1}{\beta^2}-\frac{1}{\beta_c^2} \right).
\end{align}
As one can see, the last term in \eqref{Exi}  vanishes for the conformal field, $\xi=\frac{1}{6}$, since the charge $\langle\hat{Q}_\xi\rangle$ and boundary terms $\langle\hat{B}_H\rangle$ look similar in the Rindler coordinates, but this is not the case for other static space-times due to their definition \eqref{Q}.  

For completeness, let us stress that the expectation value of the canonical Hamiltonian has the following form: 

\begin{gather}
 \langle \hat{H}_c\rangle =\langle\hat{H} \rangle+\langle \hat{B}_H \rangle
\nonumber
=\\=
\nonumber
 \int d^3 X \sqrt{g} \Bigg( \frac{\pi^2}{30} \bigg[ \frac{1}{(\beta_\mu\beta^\mu)^2}-\frac{1}{((\beta_c)_\mu(\beta_c)^\mu)^2}\bigg]
+ \frac{\pi^2}{3} \frac{1}{((\beta_c)_\mu(\beta_c)^\mu)}  \bigg[ \frac{1}{\beta_\mu\beta^\mu}-\frac{1}{(\beta_c)_\mu(\beta_c)^\mu}\bigg] \Bigg)
= \\=
A\frac{1}{2\epsilon^2}\frac{\pi^2}{30}\bigg[ \frac{1}{\beta^4}-\frac{1}{\beta_c^4}\bigg]+ A \frac{1}{2 \epsilon^2}\frac{\pi^2}{3}
\frac{1}{\beta_c^2}\bigg[ \frac{1}{\beta^2}-\frac{1}{\beta_c^2}\bigg].
\end{gather}
As one can see, the three definitions of energy are not equivalent to each other in space-times with Killing horizons.

\subsection{Euclidian path integral in Rindler coordinates}
As we have shown in the previous section, the Euclidian path integral for space-times with Killing horizons is defined as:
\begin{gather}
\label{loz3phase}
 \log Z^E =\log Z^E_3 =
 \\= -
\beta \int d^3 x \sqrt{g} \int_\infty^{m^2}d m^2  \sumint_i \frac{1}{2\omega_i}   \bigg[\triangle_3 \partial_{m^2}\phi_i(x)  \phi_i ^*(x) -  \partial_{m^2}\phi_i(x)  \triangle_3 
 \phi_i ^*(x) \bigg]n(\beta \omega_i),
\nonumber 
\end{gather}
since the sum of the first and second terms in \eqref{logZ1} vanishes: $\log Z^C_1+\log Z^C_2=0$, if the energy spectrum $\omega_i$ does not depend on the mass. 

At the same time the thermal partition function is defined as:
\begin{align}
\log Z^C=\log Tr(e^{-\beta :\hat{H}:})=- \int d^3 x \sqrt{g} g^{00} \sumint_i      \phi_i(x)  \phi_i ^*(x) \log\left(1- e^{-\beta \omega_i}\right).
\end{align} 
In the absence of horizons $\log Z_3^E$ is finite. Hence one can exchange the integrals over the momenta and coordinates to express this term in terms of scattering phase as in \eqref{phas}.  But for space-times with Killing horizon $\log Z_3^E$ is divergent. Thus, in the latter case the answer will depend on which integral is taken first, over coordinates or over momenta. 

To understand why the order of integration is important, let us consider the spatial part of the mode near the horizon \eqref{modeRindler}:
\begin{align}
\label{modeK}
 \sqrt{\frac{\omega \sinh(\pi \omega)}{\pi}} K_{i \omega}\left(\sqrt{m^2+k^2} r\right)  \approx - \sin\Big(\omega \log(\sqrt{m^2+k^2} r/2)+\gamma_\omega\Big),
\end{align}
where $\gamma_\omega$ is a phase of  $ \Gamma(1+i \omega)$. As one can see this limit works for a finite transverse momentum $k$, and the limits $k\to \infty$ and $r\to 0$ do not commute. Therefore, in such a case one can not use the definition of the Euclidean path integral in terms of phase shift \eqref{phas}, since it assumes that moving far away from the potential the phase shift should no longer depend on the distance for any momentum $k$. Hence, one should first take all momentum integrals and then take the volume integral. The resulting expression takes a clear physical meaning as the volume integral of free energy density. 

Therefore, the Euclidian path integral for a massless scalar field in the Rindler space-time has the following form:

\begin{gather}
\label{Rindler log Z}
    \log Z^E=\log Z^E_3
    =\\=  
    \nonumber 
   \int d^4 X \sqrt{g} \Bigg(   \frac{\pi^2}{90} \bigg[ \frac{1}{(\beta_\mu\beta^\mu)^2}-\frac{1}{((\beta_c)_\mu(\beta_c)^\mu)^2}\bigg]  
+ \frac{\pi^2}{9} \frac{1}{((\beta_c)_\mu(\beta_c)^\mu)}  \bigg[ \frac{1}{\beta_\mu\beta^\mu}-\frac{1}{(\beta_c)_\mu(\beta_c)^\mu}\bigg]\Bigg).
\end{gather}
Or if we introduce a volume cut-off near the horizon:
\begin{align}
    \log(Z^E)= \beta A \frac{1}{2 \epsilon^2}\frac{\pi^2}{90} \bigg[ \frac{1}{\beta^4}-\frac{1}{\beta_c^4}\bigg]  
+ \beta A \frac{1}{2 \epsilon^2}\frac{\pi^2}{9}
\frac{1}{\beta_c^2}\bigg[ \frac{1}{\beta^2}-\frac{1}{\beta_c^2}\bigg].
\end{align}
If one first takes the volume integral and then the momentum integral the result will be as follows \cite{Kabat:1995eq,Akhmedov:2021cwh,Iellici:1998np}: 
\begin{gather}
    \log Z^E=\log Z^E_3
\nonumber
    =\\=
 -A \int_\infty^{m^2}d m^2 \int_{-\infty}^\infty \frac{d^2 k}{(2\pi)^2} \int_0^\infty \frac{d \omega}{2\pi}    \log\left(1-e^{-\beta \omega}\right) \partial_{m^2} \partial_\omega \theta_{\omega},
\nonumber
= \\=
\beta A \frac{\pi^2}{3} \left[\frac{1}{\beta^2}-\frac{1}{\beta_c^2}\right] \int_{\delta^2}^\infty  \frac{d s}{(4\pi s)^{\frac{d}{2}}},
\end{gather}
where $\delta$ is an ultraviolet cutoff and $\theta_{\omega}$ is scattering phases of \eqref{modeK}. Therefore, if one first takes the volume integral in \eqref{loz3phase}, the Euclidian path integral for space-times with Killing horizons depends only on the scattering phase. This connection was pointed out recently in \cite{Law:2022zdq,Grewal:2022hlo}. In this form the value of the free energy does not have a standard form, namely $F \sim 1/\beta^d$. Due to this, in the literature, one can find different answers for the Euclidean path integral in the Rindler coordinates. For more details, see e.g. \cite{Iellici:1998np,Akhmedov:2021cwh}. 
Therefore, we consider the definition of the free energy as the volume integral of the energy as more natural.

At the same time thermal partition function $\log Z^C$ for a massless scalar field in the Rindler space-time has the following form:
\begin{gather}
\log Z^C  =  -  \int d^3 x \sqrt{g} g^{00}      \sumint_i   \phi_i(x)  \phi_i ^*(x)    \log\left(1- e^{-\beta \omega_i}\right)
=\\=
\int d^4 x \sqrt{g} \frac{\pi^2}{90}  \bigg[ \frac{1}{(\beta_\mu\beta^\mu)^2}-\frac{1}{((\beta_c)_\mu(\beta_c)^\mu)^2}\bigg].
 \nonumber
 \end{gather}

As one can see the first term in the Euclidian path integral \eqref{Rindler log Z}  coincides with the thermal partition function $\log Z^C$. However, the Euclidian path integral \eqref{Rindler log Z} contains an additional contributions, the physical meaning of which is not clear for us yet.  

Now let us look at the derivatives with respect to the inverse temperature of the Euclidean path integral:
\begin{gather}
-\partial_\beta \log(Z^E)=  A \frac{1}{2 \epsilon^2}\frac{\pi^2}{30} \bigg[ \frac{1}{\beta^4}-\frac{1}{\beta_c^4}\bigg]  
+  A \frac{1}{2 \epsilon^2}\frac{\pi^2}{9}
\frac{1}{\beta_c^2}\bigg[ \frac{1}{\beta^2}-\frac{1}{\beta_c^2}\bigg]+\text{const.},
\end{gather}
here only the first term coincides with the expectation value of the Hamiltonian. In \cite{Moretti:1997qn}, the author concludes that the derivative of the Euclidean path integral is equal to the energy \eqref{Exi} for $\xi=1/9$:
\begin{align}
    -\partial_\beta \log(Z^E) =\langle\hat{E}_{\xi=\frac{1}{9}}\rangle+const.= \langle\hat{H}\rangle+\langle\hat{Q}_{\xi=\frac{1}{9}}\rangle+\langle\hat{B}_H\rangle+const..
\end{align}
Or since the two boundary terms \eqref{QRin} and \eqref{BRin} are linearly releted in the Rindler coordinates, $  \langle\hat{B}_H\rangle=-\frac{1}{6\xi}   \langle\hat{Q}_\xi\rangle$, one can rewrite it as follows:
\begin{align}
    -\partial_\beta \log(Z^E) - \langle\hat{H}\rangle=\frac{1}{3}\langle\hat{B}_H\rangle+const..
\end{align}
But it is just a coincidence, since there is no such a dependence for higher dimensional cases, i.e.:
\begin{align}
    -\partial_\beta \log(Z^E) -\langle\hat{H}\rangle \nsim \langle\hat{B}_H\rangle+const..
\end{align}
 For more detail see the table:

\begingroup
\renewcommand*{\arraystretch}{2.2}
\begin{center}
\begin{tabular}[2]{ |c|c|c|c| } 
 \hline
  $\textbf{density of:}$   & $d=6$ & $d=8$   
  \\ 
  \hline
 $-\partial_\beta \log(Z^E)$ &  $\frac{\pi^3}{r^6}\left(\frac{2}{189 \beta^6}+\frac{1}{15\beta^4 \beta_c^2}+\frac{8}{45 \beta^2 \beta_c^4}-\frac{241}{945 \beta^6_c}\right)$ & $\frac{\pi^4}{r^8}\left( \frac{1}{225 \beta^8}+\frac{8}{189 \beta^6 \beta_c^2}+\frac{14}{75 \beta^4\beta_c^4}+\frac{16}{35 \beta^2 \beta_c^6}-\frac{3263}{4725 \beta_c^8}\right)$
  \\
  \hline 
 $ \langle :\hat{H}:  \rangle$ & $\frac{\pi^3}{r^6}\left(\frac{2}{189 \beta^6}+\frac{1}{45 \beta^4 \beta_c^2}-\frac{52}{945 \beta^6_c}\right)$ & $\frac{\pi^4}{r^8}\left( \frac{1}{225 \beta^8}+\frac{4}{189 \beta^6 \beta_c^2}+\frac{8}{225 \beta^4\beta_c^4}-\frac{289}{4725 \beta^6_c}\right)$
   \\
  \hline 
 $-\partial_\beta \log(Z^E) - \langle H \rangle$ &   $\frac{\pi^3}{r^6}\left(\frac{2}{45\beta^4 \beta_c^2}+\frac{8}{45 \beta^2 \beta_c^4}-\frac{2}{9 \beta^6_c}\right)$ & $\frac{\pi^4}{r^8}\left( \frac{4}{189 \beta^6 \beta_c^2}+\frac{34}{225 \beta^4\beta_c^4}+\frac{16}{35 \beta^2 \beta_c^6}-\frac{2975}{4725 \beta_c^8}\right)$
   \\
  \hline
  $\langle\hat{B}_H\rangle$ & $\frac{\pi^3}{r^6}\left(\frac{4}{45\beta^4 \beta_c^2}+\frac{8}{9 \beta^2 \beta_c^4}-\frac{44}{45 \beta^6_c}\right)$ & $\frac{\pi^4}{r^8}\left( \frac{4}{105 \beta^6 \beta_c^2}+\frac{2}{5 \beta^4\beta_c^4}+\frac{16}{5 \beta^2 \beta_c^6}-\frac{382}{105 \beta_c^8}\right)$
   \\
\hline
  $\langle\hat{Q}_\xi\rangle$ & $-5\xi \langle\hat{B}_H\rangle $ & $-\frac{14}{3}\xi \langle\hat{B}_H\rangle $
   \\
  \hline 
\end{tabular}
\end{center}

In any case, if one uses the standard method then the fundamental statistical-mechanical relation does not hold. 

At the same time the new method of computation of the Euclidean path integral proposed in the Section \ref{sec3} gives:
\begin{align}
\log Z^E=\lim_{M\to 0}\log Z^E_M=-\lim_{M\to 0}\frac{\beta}{2}\int d^3 x\sqrt{-g} g^{00}\int_\infty^{M^2} d M^2  \ G_M (x,x),
\end{align}
where the Feynman thermal propagator at the coincident points with the new mass term has the following form: 
\begin{align}
    G_M (x,x) =\frac{1}{\beta}\sum_{\omega_n}\int \frac{d^2 \vec{k}}{(2 \pi)^2} \int_0^{\infty} \frac{ d \omega}{\pi^2} \frac{2\omega\sinh\pi \omega}{\w_n^2+\omega^2+M^2}K_{i \omega}\big(kr\big)K_{i\omega}\big(k r\big).
\end{align}
As a result one obtains: 
\begin{align}
\log Z^E = \frac{\pi^2}{90} \int d^4 x \sqrt{g}  \bigg[ \frac{1}{(\beta_\mu\beta^\mu)^2}-\frac{1}{((\beta_c)_\mu(\beta_c)^\mu)^2}\bigg].
\end{align}

Therefore, the Euclidian path integral has the correct form, since its derivative with respect to the inverse temperature coincides with the expectation value of the Hamiltonian. Let us also stress that this result is in agreement with the WKB result from \cite{Susskind:1994sm} .  Furthermore, this results is generally covariant and it looks like the coordinate transformation of the standard Minkowskian thermal partition function. At the same time the second term in \eqref{Rindler log Z} does not fit into this reasoning.

\section{Conclusion, discussion and questions}
In this paper we check whether the Euclidean path integral equals the thermal partition function for the non-minimally coupled scalar quantum field theory or not. We make the following observations:
\begin{itemize}
    \item  For spaces with compact spatial section the main result is that the Euclidean path integral $Z^E$ is equal to the thermal partition function of the canonical ensemble $Z^C$.  This case is widely discussed in the literature.
\item  If spatial sections are non-compact and there is no Killing horizon, then the standard method implies that the equality between the Euclidean path integral and thermal partition function approaches holds, since two additional terms in \eqref{logZm}, which are sensitive to the scattering phase for the spatial part of the mode, cancel each others. 
\item If spatial sections are bordered by Killing horizons, then the situation is quite different from the last one. First, space-times with Killing horizons are endowed with their own temperature due to the Hawking type radiation. Second, the single particle energy spectrum of the theory does not depend on the mass, which leads to the fact that the standard method of computation of the Euclidean path integral is defined only by the third term in \eqref{logZm}. This fact, generally speaking, does not lead to the equality between the Euclidean path integral and thermal partition function approaches. Third, there are three different definitions of the energy, and as we have shown, the derivative with respect to the inverse temperature does not give the value of any of these three types of energy.  
\item We propose a new method of computation the Euclidean path integral. This method gives a self-consistent answer for compact cases and gives the correct equality between the Euclidean path integral and thermal partition function approaches for non-compact cases with and without Killing horizons. 
\end{itemize}

Let us also provide a list of questions that we would like to address in the future work:

\begin{itemize}
    \item In the Section \ref{sec2}, we have shown that the thermal partition function is not generally covariant, since it can be represented as the Euclidean path integral with a non-covariant measure, but with a standard action. Or it can be represented as the Euclidean path integral with a covariant measure, but with an action with an optical metric (obtained by a conformal transformation from the original one). But, as we have seen, for compact spaces and non-compact spaces without Killing horizons the thermal partition function is equal to the Euclidean path integral, which is generally covariant. Thus the questions is that is the thermal partition function is generally covariant for non-compact manifolds with Killing horizons? 
    \item In this article we show that in space-times with Killing horizons in general the fundamental statistical-mechanical relation does not hold. However, in \cite{Dowker:2010bu} the related is used to obtain conformal anomaly for the de Sitter space-time. The key point is that the same result is obtained as in \cite{Fursaev:1993hm}, where the author uses the zeta function method to compute the Euclidean path integral. So the question is as follows. Does the fundamental statistical-mechanical relation hold in the presence for the logarithmic term in general? 
    \item Recently, Zubarev's approach to the description of thermodynamics has been discussed in the literature (for more detail see: \cite{
Prokhorov:2018bql, Becattini:2017ljh, Becattini:2019dxo, Becattini:2019poj, Prokhorov:2019yft}). This method appears to be generally covariant and the partition function is built with the ise of the stress energy tensor: $Z^C_T=Tr\left[exp\left(\int d \Sigma^\mu \hat{T}_{\mu \nu} \beta^\nu\right)\right]$. Since, the Energy defined by the stress-energy tensor ($E$) is different from the energy determined by the Hamiltonian (which stands in the definition of the thermal partition function) in space-time with Killing horizons. It is important to understand, which of the definitions of the partition function is correct? 
\end{itemize}

\section*{Acknowledgments}
We would like to acknowledge valuable discussions with K.V.Bazarov, D.I. Sadekov and D.A. Trunin. Especially we would like to thank E.T.Akhmedov for valuable discussions, sharing his ideas and correcting the text. This work was supported by the grant from the Foundation for the Advancement of Theoretical Physics and Mathematics ``BASIS'', by the Euler grant from the Saint Petersburg Leonhard Euler International Mathematical Institute and supported by the Ministry of Science and Higher Education of the Russian Federation (agreement no. 075–15–2022–287).

\begin{appendices} 

\setcounter{equation}{0}
\renewcommand\theequation{A.\arabic{equation}}

\section{Thermal Feynman propagator \label{appA} }
\numberwithin{equation}{section}
Here we will show that in the static space-time there are two different representations of  the thermal Feynman propagator. 

The first method is based on the eigen-functions of the Klein-Gordon operator:
\begin{align}
     \left(-\Box_E+m^2+\xi R \right)\psi_i(x)=\lambda_i \psi_i(x),
\end{align}
which are orthonormal and complete:
\begin{align}
\int d^4x \sqrt{g} \psi^*_i(x) \psi_j(x)=\delta_{ij} \quad \text{and}\quad \sumint_i \psi^*_i(x) \psi_i(y) =\frac{\delta(x-y)}{\sqrt{g}}.
\end{align}
Hence the Green function for the Klein-Gordon equation:
\begin{align}
\label{KleinN}
    \left(-\Box_E+m^2+\xi R \right) G(x,y)=\frac{\delta(x-y)}{\sqrt{g}}
\end{align}
can be represented as:
\begin{align}
    G(x,y)=\sum_i \frac{\psi^*_i(x) \psi_i(y)}{\lambda_i}.
\end{align}
The second method is based on the eigen-functions of the following operator:
\begin{align}
   ( -\partial_\tau^2+H_s^2)\chi_i(x)=g_{00}\left(-\Box_E+m^2+\xi R \right) \chi_i(x)= \rho_i \chi_i(x),
\end{align}
which are orthonormal and complete:
\begin{align}
\int d^4x \sqrt{g} g^{00}\chi^*_i(x) \chi_j(x)=\delta_{ij} \quad \text{and}\quad \sumint_i \chi^*_i(x) \chi_i(y) =\frac{\delta(x-y)}{\sqrt{g}g^{00}}.
\end{align} 
Hence, the corresponding Green function:
\begin{align}
\label{G2}
     G(x,y)=\sumint_i \frac{\chi^*_i(x) \chi_i(y)}{\rho_i},
\end{align}
solves the same Klein-Gordon equation\eqref{KleinN}. Therefore, two methods lead to the same Green function, since they solve the same equation. Note that we are working in Euclidean signature rather than Lorentzian one.

The third method is based on the canonical quantization. The Klein-Gordon equation in Lorentzian
signature is:
\begin{align}
\label{eqmotion}
(g^{00}\partial_t^2-\triangle_3  +m^2+\xi R) \varphi=0.
\end{align}
where $\triangle_3=g^{-\frac{1}{2}}\partial_i\left(g^{i j} g^{\frac{1}{2}}\partial_j\right) $. The solution of this equation of motion has the following form:
\begin{align}
\hat{\varphi}= \sumint_i \left[ \frac{e^{-i \omega_i t} }{\sqrt{2 \omega_i}} \phi_i(x) \hat{a}_{i}+h.c. \right],
\end{align}
where $\phi_i(x)$ is the spatial part of the mode, that solves:
\begin{align}
\label{eqrefmode}
    (-g^{00}\omega_i^2-\triangle_3  +m^2+\xi R) \phi_i(x)=0.
\end{align}
Using the canonical commutation relations:
\begin{align}
\left[\hat{\varphi}(x), \dot{\hat{\varphi}}(y)\right]=i \frac{\delta(x-y)}{\sqrt{g}g^{00}},
\end{align}
one can obtain the completeness and orthogonality relations for the spatial part of the modes:

\begin{align}
\label{ort}
\int d^3 x\sqrt{g} g^{00} \phi_i(x)\phi^*_j(x) =\delta_{i,j} \quad \text{and }\quad \sumint_i \phi_i(x)\phi^*_i(y)=\frac{\delta^3(x-y)}{\sqrt{g}g^{00}}.
\end{align}
The Hamiltonian of the theory is defined as follows:
\begin{gather}
\label{t}
:\hat{H}:
=\\=
\nonumber
- :\int d^3 x \sqrt{g} g^{00} \left( \partial_t \hat{\varphi}(x)\partial_t \hat{\varphi}(x)-\frac{g_{00}}{2}\left[g^{00}\partial_t \hat{\varphi}(x)\partial_t \hat{\varphi}(x) +\triangledown_i\hat{\varphi}(x)\triangledown^i\hat{\varphi}(x)+m^2\hat{\varphi}(x)^2 +\xi R  \hat{\varphi}(x)^2 \right]   \right):
=\\=
\nonumber
: \int d^3 x \sqrt{g} g^{00} \frac{1}{2}  \left(  -\partial_t \hat{\varphi}(x)\partial_t \hat{\varphi}(x)+\hat{\varphi}(x)\partial^2_t\hat{\varphi}(x)    \right):+\frac{1}{4}\int d^3 x \partial_i[g^{i j}\sqrt{g}\partial_j \hat{\varphi}^2(x)]
=\\=
\nonumber
\sum_i  \ \omega_i \ \hat{a}^\dagger_{i} \hat{a}_{i},
\end{gather}
where we use the equation of motion \eqref{eqrefmode} and completeness relation \eqref{ort}.  The key point here is that the following boundary contribution vanishes:
\begin{align}
\label{boundHam}
   \frac{1}{4} \int d^3 x \partial_i[g^{i j}\sqrt{g}\partial_j \hat{\varphi}^2(x)]=0.
\end{align}
But this is not the case for space-times with Killing horizons. 

Using the Bose-Einstein or Planckian distribution:
\begin{align}
\langle\hat{a}^\dagger_{i} \hat{a}_{j}\rangle_\beta=\delta_{ij}  n(\beta \omega_i), \quad\text{where} \quad n(\beta \omega_i)= \frac{1}{e^{\beta \omega_i}-1},
\end{align}
one can obtain the thermal Wightman function:
\begin{gather}
W(t,x_1,x_2)
= \\=
\nonumber
\sumint_i \frac{e^{-i \omega_i  t }}{2 \omega_i}  \phi_i(x_1)  \phi_i^*(x_2)(1+ n(\beta \omega_i)) 
+
\sumint_i \frac{e^{i \omega_i  t }}{2 \omega_i}  \phi_i^*(x_1)  \phi_i(x_2) n(\beta \omega_i).
\end{gather}
Using this Wightman function, one can construct the thermal Feynman propagator as follows:
\begin{align}
\label{Gn}
G(t,x_1,x_2)=\theta(t)W(t,x_1,x_2)+\theta(-t)W(-t,x_2,x_1),
\end{align}
which after the analytical continuation to the Euclidean signature acquires the following form:
\begin{align}
    G(x,y)=\sumint_{n,i} \frac{\beta^{-1/2}e^{-i w_n  t_1 } \phi_i(x_1) \left[\beta^{-1/2} e^{-i w_n  t_2 }  \phi_i(x_2) \right]^* }{w_n^2+\omega_i^2},
\end{align}
where $w_n =\frac{2\pi n}{\beta}$ are the Matsubara frequencies. Such a representation is in full agreement with \eqref{G2}. Therefore, if one uses the canonical commutation relations to quantize the field, and then constructs the thermal Feynman propagator, one will obtain the same expression as in the second method.

Let us also stress that if one uses the optical metric, then the all method give the same representation for the propagator, since $g_{00}=1$. 
\end{appendices}
\newpage

\bibliographystyle{unsrturl}
\bibliography{bibliography.bib}

\end{document}